\documentstyle[11pt]{article}

\newcommand{\noi}{\noindent}
\newcommand{\bc}{\begin{center}}
\newcommand{\ec}{\end{center}}

\def\ifmath#1{\relax\ifmmode #1\else $#1$\fi}
\def\3quarter{{\textstyle{3 \over 4}}}

\def\ra{\rightarrow}

\def\lf{\leaders\hbox to 1em{\hss.\hss}\hfill}

\def\21{$SU(2) \ot U(1)$}
\def\ne{\hbox{$\nu_e$ }}
\def\nm{\hbox{$\nu_\mu$ }}
\def\nt{\hbox{$\nu_\tau$ }}

\def\ns{\hbox{$\nu_S$ }}

\def\etal{\hbox{\it et al., }}
\def\J.W.F.V{\hbox{J. W. F. Valle }}

\def\sym{\hbox{symmetry }}
\def\sym{\hbox{symmetry }}

\def\gau{\hbox{gauge }}

\def\neu{\hbox{neutrino }}
\def\sa{\hbox{such as }}

\def\neus{\hbox{neutrinos }}

\def\Phys{\hbox{Physics }}

\def\neup{\hbox{neutrino. }}

\def\eq#1{{eq. (\ref{#1})}}

\def\VEV#1{\left\langle #1\right\rangle}

\def\lsim{\raise0.3ex\hbox{$\;<$\kern-0.75em\raise-1.1ex\hbox{$\sim\;$}}}
\def\gsim{\raise0.3ex\hbox{$\;>$\kern-0.75em\raise-1.1ex\hbox{$\sim\;$}}}

\def\bel{\begin{letter}}
\def\eel{\end{letter}}
\def\beq{\begin{equation}}
\def\eeq{\end{equation}}
\def\bef{\begin{figure}}
\def\eef{\end{figure}}
\def\bet{\begin{table}}
\def\eet{\end{table}}
\def\bea{\begin{eqnarray}}
\def\ba{\begin{array}}
\def\ea{\end{array}}
\def\bi{\begin{itemize}}
\def\ei{\end{itemize}}
\def\ben{\begin{enumerate}}
\def\een{\end{enumerate}}
\def\ra{\rightarrow}
\def\ot{\otimes}

\def\eea{\end{eqnarray}}

\def\np#1#2#3{           {\it Nucl. Phys. }{\bf #1} (19#2) #3}
\def\pl#1#2#3{           {\it Phys. Lett. }{\bf #1} (19#2) #3}
\def\pr#1#2#3{           {\it Phys. Rev. }{\bf #1} (19#2) #3}

\def\rmp#1#2#3{          {\it Rev. Mod. Phys. }{\bf #1} (19#2) #3}

\def\n.c.#1#2#3{         {\it Nuovo Cim. }{\bf #1} (19#2) #3}
\def\r.n.c.#1#2#3{       {\it Riv. del Nuovo Cim. }{\bf #1} (19#2) #3}

\relax

\def\lsim{\raise0.3ex\hbox{$\;<$\kern-0.75em\raise-1.1ex\hbox{$\sim\;$}}}
\def\gsim{\raise0.3ex\hbox{$\;>$\kern-0.75em\raise-1.1ex\hbox{$\sim\;$}}}
\newcommand {\unit}[1]{\hspace{0.33em} {\rm #1}}
\newcommand {\ignore}[1]{}

\begin{document}
\begin{titlepage}
\noi
\today
\begin{center}
\hfill FTUV/93-04\\
\hfill IFIC/93-04\\
\hfill Submitted to Nuclear Physics B\\
\vskip 1cm
{\LARGE \bf Reconciling Dark Matter, Solar and Atmospheric
Neutrinos}\\
\vskip 0.5cm
{\LARGE J. T. Peltoniemi}
\footnote{E-mail 16444::PELTONIE or PELTONIE@EVALVX}
\vskip 0.2cm
{\LARGE J. W. F. Valle}
\footnote{E-mail 16444::VALLE or VALLE at EVALUN11}
\vskip .2cm
{\it Instituto de F\'{\i}sica Corpuscular - C.S.I.C.\\
Departament de F\'{\i}sica Te\`orica, Universitat de Val\`encia\\
46100 Burjassot, Val\`encia, SPAIN}\\
\vskip .5cm
{\bf Abstract}
\end{center}
\begin{quotation}
We present models that can reconcile the solar
and atmospheric neutrino data with the existence of
a hot dark matter  component in the universe. This
dark matter is a quasi-Dirac neutrino whose mass
$m_{DM}$ arises at the one-loop level. The solar
\neu deficit is explained via nonadiabatic
conversions of \ne to a sterile \neu $\nu_s$
and the atmospheric \neu data via maximal \nm
to \nt oscillations generated by higher order
loop diagrams.
%
For $m_{DM} \sim 30$ eV the radiative
neutrino decay can lead to photons that can
ionize interstellar hydrogen.
In one of the models one can have observable
$\nu_e$ to $\nu_\tau$ oscillation rates, with
no appreciable \nm oscillations at accelerator
experiments. In addition, there can be observable
rates for tau number violating processes \sa
$\tau \ra 3e$ and $\tau \ra e + \gamma$.
In the other model one can have sizeable
$\nu_e$ to $\nu_\mu$ oscillation rates, as
well as sizeable rates for muon number violating
processes \sa $\mu \ra e + \gamma$,
$\mu \ra e + majoron$ and $\mu \ra 3e$.

\end{quotation}
\vfill
\end{titlepage}

\section{Introduction}

The existing hints in favour of nonzero \neu masses
include the existence of a hot dark matter component
as recently suggested by COBE data and the deficits
observed in the solar \ne and atmospheric \nm fluxes
\cite{cobe,Davis,GALLEX1,SAGE,KAMII,atm}. Neutrino
oscillations would provide the most attractive
explanation of these fluxes, which seem to be
in conflict with standard theoretical
expectations \cite{bp,Gaisser}.
A common understanding of these data
seems problematic, even at the level of simultaneously
fitting the three phenomena. For example, one could have
$\nu_e \ra \nu_\mu$ oscillations in the sun with
$\delta m^2 \sim 10^{-5} \rm eV^2$, with the $\nu_\tau$ as
the hot dark matter (HDM) component ($m_{\nu_\tau} \sim \rm{few} $~eV).
However, in this scenario there is no room for oscillations to account
for the atmospheric neutrino anomaly.
Another possibility is again to have $\nu_e \ra \nu_\mu$
oscillations in the sun, while $\nu_\mu \ra \nu_\tau$
oscillations are responsible for the understanding of
the atmospheric neutrino data. In this case there is
no room for hot dark matter.

This suggests the need for the existence of an
additional light \neu state, which must be
sterile in order not to affect the invisible
Z width, successfully predicted in the standard
model.

In this letter we propose an alternative scenario
that includes such a sterile neutrino $\nu_s$. In this
model the neutrino mass scales required for the
joint explanation of the HDM, the solar and
the atmospheric \neu data arise
from radiative corrections associated to new
Higgs bosons at the electroweak scale. More
than just a fit, our model provides an
underlying theory that successfully
explains the origin of these scales. Unlike
the case of seesaw models \cite{Bludman92}, we relate the
small $m_{\odot}/m_{DM}$ ratio to a quantum
mechanical loop suppression factor.
The dark matter mass parameter $m_{DM}$
arises at the one-loop level, while the
oscillations responsible for the explanation
of the solar and atmospheric \neu deficits
arise only at two and three loops.

In our model the HDM consists of a quasi-Dirac neutrino made mainly of
\nt and $\nu_\mu$. One would get the best fit to the observations by
choosing its mass to be about 3 eV leading to
$\Omega_\nu \sim 0.3$ \cite{Tayloretal}. In fact, a 3 eV
Dirac-neutrino might fit the observations even better than the usually
considered 7 eV Majorana-neutrino, as it would give more power on the largest
scales.
On the other hand,
if the HDM mass is chosen as $m_{DM} \sim 30$ eV
the photons produced in the radiative decays,
$\nu_{DM} \ra \nu + \gamma$ have just the right
properties required in order to ionize
interstellar hydrogen, as suggested by
observation \cite{Sciama91}.
The solar neutrino data are explained via the MSW
effect involving nonadiabatic $\nu_{e} \ra \nu_{S}$
transitions, while the atmospheric neutrino data can be explained
via $\nu_{\mu} \ra \nu_{\tau}$ oscillations with maximal
mixing and $\delta M^2 \sim 10^{-2} - 10^{-3} {\rm eV}^2$.
These values are in agreement with the data of
the Kamiokande group and with some of the IMB
studies, although not those involving stopped
muons \cite{atm}.

There can also be observable $\nu_e \ra \nu_\tau$
oscillation rates, that could be probed as a
byproduct of \nm to \nt searches at the new
generation of \neu oscillation experiments
\cite{chorus}. There are, however, no appreciable
\nm to \ne or \nm to \nt oscillations on scales
that can be presently probed by accelerator experiments.
The model also leads to tau number violating processes
\sa $\tau \ra 3e$ and $\tau \ra e + \gamma$ with rates
that can lie within the experimental sensitivities of
the next generation of experiments.

In an alternative model based on the
$e-\tau+\mu$ symmetry the experimentally
observable oscillation rates are for
$\nu_e \ra \nu_\mu$, and similarly
lepton flavour violating processes
now are $\mu \ra e + \gamma$, $\mu \ra 3e$ and $\mu \ra e + J$
($J$ denotes  the majoron), with rates that can lie within the
present experimental sensitivities.

Both models are compatible with laboratory,
astrophysical and cosmological observations,
including primordial nucleosynthesis limits.

\section{Neutrino Masses and Mixing}

The existing hints from HDM, solar and atmospheric
\neu observations restrict the form of the \neu
mass and mixing. As discussed above, in order
to provide a common interpretation of these data
one requires the existence of a sterile \neup
If we restrict ourselves to matrices that
have a simple underlying symmetry, we have
two interesting possibilities to consider: either
we choose the sterile \neu to be at the
HDM scale or at the MSW scale. The
first possibility was considered in
ref. \cite{DARK92}. Here we focus
on the alternative logical possibility
that the sterile \neu is at the solar \neu mass scale.
We also require the existence of a symmetry
defined so that the HDM scale is
invariant, while the solar and atmospheric
\neu oscillations arise as breaking effects.
Our strategy follows closely the lines of
ref. \cite{smi2} and \cite{juha}.

We may write
the neutrino mass matrix in the following form
\beq
M_\nu = \left(
\begin{array}{llllll}
\mu & a_e & a_\mu & a_\tau \\
a_e & \epsilon_{ee} & m & \epsilon_{e \tau} \\
a_\mu & m & \epsilon_{\mu\mu} & M \\
a_\tau & \epsilon_{e \tau} & M & \epsilon_{\tau\tau} \;
\end{array}
\right)
\label{nino}
\eeq
where the basis is ($\nu_s,\nu_e,\nu_\mu,\nu_\tau$)
\footnote{Another phenomenologically viable
alternative, considered in section 6,
corresponds to having the same matrix
but in the reversed basis ($\nu_s,\nu_e,\nu_\tau,\nu_\mu$).
}.
For values of the entries $m,M \gg \mu, a_i, \epsilon_{ij}$
this matrix has an approximate $e-\mu+\tau$ symmetry.

The \neu mass eigenstates are given in terms of the weak
eigenstates as
\begin{eqnarray}
\nu_1 &\approx& \cos \theta_m \cos \theta \nu_e - \cos\theta_m \sin\theta
\nu_\tau + \sin\theta_m \nu_s\\
\nu_2 &\approx& -\sin \theta_m \cos \theta \nu_e + \sin\theta_m \sin\theta
\nu_\tau + \cos\theta \nu_s \\
\nu_3 &\approx& \frac{1 }{\sqrt{2} } \cos\theta \nu_\tau + \frac{1 }{\sqrt{2}
} \sin\theta \nu_e  + \frac{ 1}{\sqrt{2} } \nu_\mu \\
\nu_4 &\approx& \frac{1 }{\sqrt{2} } \cos\theta \nu_\tau + \frac{1 }{\sqrt{2}
} \sin\theta \nu_e  - \frac{ 1}{\sqrt{2} } \nu_\mu
\end{eqnarray}
where now $\theta$ denotes the mixing between $\nu_e$ and $\nu_\tau$, and
$\theta_m$ is the mixing of the lightest states relevant for
the solar \neu deficit.

\sloppy
For suitable choices of parameters obeying this
hierarchy the heaviest \neu is quasi-Dirac, formed by \nm
and \nt and its mass $m_{DM}$ can be chosen to be at
the HDM scale. The remaining \neus have much smaller
masses at the MSW scale, and their mixing can explain the
deficit of solar \ne neutrinos. Finally, the
splitting between \nm and \nt generates
oscillations that can explain the observed \nm
deficit in the atmospheric neutrino flux.

Neutrino oscillations are characterized by three
different oscillation lengths. The shortest one is
due to the mass squared difference between the heavier and
lighter states, $m_{DM}^2$, while the two long-scale oscillations are
due to the squared mass difference of the lightest states $\delta m^2$,
responsible for the explanation of the solar \neu data,
and that of the heaviest
states, denoted by $\delta M^2$, chosen to fit
the atmospheric \neu data.

For typical energies, the oscillations at scale $m_{DM}^2$
have oscillation lengths that can lie in the
region of sensitivity of accelerator experiments and
their rates could be observable. The only significant oscillation
at this level occurs between electron and tau neutrinos,
and is characterized by a mixing angle $\theta$.
The oscillations of muon neutrinos remain
experimentally unobservable in accelerator
based experiments. 

The oscillations to the sterile neutrino \ns are too
small to be observed in laboratory, but they could
lead to important effects in astrophysics and cosmology. Indeed, \ne,
\nm or \nt conversions could populate the sterile
states \ns in the early universe thus increasing
the effective light neutrino number and consequently
the primordial light element abundances. In order to prevent this the
corresponding conversion probabilities must be
smaller than $3.4 \times 10^{-6} {\rm eV^2}$
\cite{dolgov}, where we have adopted the most stringent
and updated nucleosynthesis limit $\Delta N_\nu < 0.3$.
This would transform into a limit for the mass entries,
\begin{equation}
a_e, a_\tau \cos\theta + a_e \sin\theta \lsim 0.002 \unit{eV}.
\end{equation}
Of all the oscillations $\nu_i \ra \ns$ with solar \neu scale
$\delta m^2$, the only one that could be important
once the above limits are fulfilled is from \ne into \ns.
This is precisely the channel that is responsible for
the explanation of the solar \neu data by the MSW effect
and is characterized by the mixing angle $\theta_m$.
The nucleosynthesis bound implies that these solar
\neu conversions must take place in the non adiabatic
regime, excluding the large mixing solution.

The oscillation from $\nu_\mu$ to $\nu_\tau$ is
important for the explanation of atmospheric neutrinos.
Due to the assumed symmetry structure, this
oscillation is characterized by almost maximal mixing.
In order to account for the observed deficit of muon neutrinos,
one should have $\delta M^2 \sim 10^{-2} - 10^{-3} {\rm eV}^2$.

\section{A Model}

We now briefly describe a model that naturally
embodies the structure described above, allowing
a simple explanation of the HDM, solar and atmospheric
\neu puzzles. The lagrangean is given by
\begin{eqnarray}
\label{Yuk}
\sum_{i,j} h_{ij} {\bar{\Psi}}_{Li} \phi \ell_{Ri} \:+ \:
f_{e\mu} \Psi_e^T C i \tau_2 \Psi_\mu s^+ \:+ \:
f_{\mu\tau} \Psi_\mu^T C i \tau_2 \Psi_\tau s^+ \:+ \:
\nonumber \\
\xi_e \bar{e}_R \nu_s \eta^- \:+ \:
\xi_\tau \bar{\tau}_R \nu_s \eta^- \:+ \:
\sum_{\ell, \ell'= e,\tau} g_{\ell \ell'} \ell_R^T C {\ell_R}' \kappa^{++} \:+
\nonumber \\
M_{s\phi \phi} \phi^T \tau_2 \tilde{\phi} s^+ \:+ \:
M_{s\eta \kappa}  {s^+} \eta^{+} \kappa^{--} \:\:+\:
M_{\eta \eta \Xi} \eta^{+} \eta^{+} \Xi^{--} \:\:+\:
M_{\kappa \Xi}^2 \Xi^{--} \kappa^{++} \:+ \:
\rm h.c.
\end{eqnarray}
The quantum numbers are summarized in the table.
The terms on the first line in the above equation
generate the entries m and M in \eq{nino} at the
one-loop level, as a result of electroweak breaking
(see Fig 1a) and may be written as \cite{BertoliniSantamaria,BabuMathur}
\beq
M \simeq m_{DM} \simeq \frac{ m_\tau^2 f_{\mu \tau} g_2}{32\pi^2 M_W}
\frac{\tilde{v}}{v}
\left(\frac{\ln\left(\frac{m_\tau^2}{M_1^2}\right)}{1-\frac{m_\tau^2}{M_1^2}}
-
\frac{\ln\left(\frac{m_\tau^2}{M_2^2}\right)}{1-\frac{m_\tau^2}{M_2^2}}
\right)
\label{M}
\eeq
where $g_2$ is the $SU(2)$ \gau coupling, $M_1$ and $M_2$ are
the physical masses of the relevant singly charged Higgses,
their mixing angle is denoted $\beta$, and $v$ and
$\tilde{v}$ denotes the vacuum expectation values
of the two Higgs doublets. These parameters can
easily be chosen so as to provide the mass of the HDM neutrino.
The mixing of this dark matter neutrino
(see below) is determined from
\beq
\tan \theta \simeq \frac{m}{M} \simeq
\frac{m_\mu^2}{m_\tau^2}\frac{f_{\mu e}}{f_{\mu \tau}}
\label{teta}
\eeq

In the absence of the three latter terms on
the last line of \eq{Yuk} the model
would have an exact global symmetry corresponding
to $e-\mu+\tau$ conservation. We can imagine that either these
are small soft breaking terms or that, alternatively,
a new singlet field $\sigma$ with the appropriate charge
is introduced. In the latter case the symmetry would be
broken spontaneously by the vacuum expectation value
$\VEV{\sigma}$ thus generating a majoron given by
$J = {\rm Im}\; \sigma$. The spontaneous violation
of this global symmetry at or below the electroweak
scale is not in conflict with existing observations, including
astrophysics and cosmology. For example, the coupling
of the majoron to electrons does not lead excessive
stellar energy loss \cite{KIM}. Moreover, the decay
$\nu_{DM} \ra \nu_{light} + J$ can be arranged
so that the resulting lifetime is longer than
the age of the universe $\tau \sim 10^{17} $ sec.
This ensures that the HDM neutrinos are still there to
form the dark matter.

Whichever way this symmetry is broken, the other
entries in \eq{nino} will be generated by higher
order, two and three loop diagrams.
For definiteness we will assume
in what follows that the global G symmetry is broken
explicitly by soft terms quadratic and cubic, as in \eq{Yuk}.
As a result these
entries are naturally small. In addition, they have
definite charges under the G symmetry given by:
$\mu \ra -1$,
$a_e , a_\tau  \ra \frac{1}{2}$,
$a_\mu \ra -\frac{3}{2}$,
$\epsilon_{ee}, \epsilon_{e \tau}, \epsilon_{\tau\tau} \ra 2$
and $\epsilon_{\mu\mu} \ra -2$.
In contrast, the entries $m$ and $M$ are invariant
under the $U(1)_G$ symmetry.

The values of the small mass matrix elements
$a_i,\mu$ may be obtained from the two-loop
diagrams in Fig. 1, while the $\epsilon_{ij}$ arise
only at three-loop level. They may be estimated as
\begin{equation}
        \mu \sim \sum_{\alpha \beta = e \tau}
        \frac{\xi_\alpha \xi_\beta g_{\alpha \beta} M_{\eta\eta \Xi}
M_{\kappa \Xi}^2
         }
        {256 \pi^4 M_0^2}
\end{equation}
\begin{equation}
        a_\ell \sim \sum_{\alpha  = e \tau}
        \frac{m_\ell \xi_\alpha g_{\ell \alpha } M_{s\phi\phi}
        M_{s\eta\kappa}}
        {128 \pi^4 M_0^2 }
        \frac{\tilde{v}}{v}
\label{a1}
\end{equation}
\begin{equation}
        a_\mu \sim \sum_{\alpha  = e \tau}
        \frac{m_\alpha f_{\mu \alpha}
         \xi_\alpha M_{s\eta\kappa} M_{\eta\eta\Xi} M_{\kappa \Xi}^2
         }
        {512 \pi^4 M_0^4}
\label{a2}
\end{equation}
\begin{equation}
\epsilon_{\ell \ell'} \sim \frac{\sin^2 2\beta
m_\ell m_\ell' g_{\ell\ell'}
M_{s\eta\kappa}^2 M_{\eta\eta\Xi} M_{\kappa\Xi}^2 }
{8192 \pi^6 M_0^4 \tilde{v}^2
}
\end{equation}
\begin{equation}
\epsilon_{\mu\mu} \sim  \sum_{\alpha \beta} \frac{m_\mu^2  f_{\mu \alpha}
f_{\mu\beta} g_{\alpha \beta}
M_{s\eta\kappa}^2 M_{\eta\eta\Xi} M_{\kappa\Xi}^2 }
{8192 \pi^6 M_0^6  }
\end{equation}
where $\ell = e, \tau$, and
$M_0$ is a typical scalar mass.

It is easy to choose the parameters so that
they can account for the dark matter and solar neutrino
masses and mixing ($\delta m^2 \sim 3\times 10^{-6} \rm eV^2$, $\sin^2
2 \theta_m \sim 0.007$), while $\delta M^2$ lies
in the range suggested by some atmospheric neutrino
observations ($10^{-3} - 10^{-2} \unit{eV}^2$).
For example,  one could choose
$f_{e\mu}= 10^{-3} $, $f_{\mu\tau}= 10^{-4} $ , $g_{ee}=0.05 $,
$g_{\tau e} = 0.005 $, $g_{\tau\tau}=0.07$,
$\xi_e=4\times 10^{-4} $, $\xi_\tau= 2 \times 10^{-5}$,
$M_{s\phi\tilde{\phi}}=0.1 $
GeV,
$ M_{\kappa\Xi}=3$ GeV, $M_{s\eta\kappa}= M_{\eta\eta\Xi}=120$
GeV,
$M_0=45 $ GeV, $v = 200$ GeV, $\tilde{v}=130$ GeV, $\sin \beta = 0.3$.

Note that it is also possible to obtain a
variant model by exchanging the signs of
the $U(1)_G$ charge assignment for $\nu_s$,
$\eta$, $\kappa$ and $\Xi$. If the scalar
$\sigma$ were present, its G assignment
would also be reversed in this new model.
In what follows we will, for simplicity,
not consider this variant.

\section{Magnetic Moments and Radiative Decay}

The transition magnetic moment matrix of neutrinos is
generated by graphs similar to those that generate
the mass matrix, but with a photon line inserted.
The entry connecting electron and muon neutrinos
is given by \cite{BabuMathur}
\beq
\mu_{\nu_\mu \nu_e}  \simeq \frac{e m_\mu^2 f_{\mu e} g_2}{16\pi^2 M_W}
\frac{\tilde{v}}{v}
\sin 2\beta \left[ \frac{1-r^2 }{r^2 } \left( 1- \ln\frac{M_1^2}{m_\mu^2}
\right) - \frac{1 }{r^2 } \ln r^2 \right]
\label{Muu}
\eeq
where $r=M_2/M_1$.

Although the individual components of
the transition magnetic moment matrix are approximately
proportional to the same entries in the mass matrix,
the full matrices are not. Indeed, on very general grounds
one knows that the mass matrix of two-component neutrinos
is symmetric, while the magnetic moment matrix is
antisymmetric \cite{BFD}.
Moreover the form factors are different and
it is easy to check that, after suitable
rotation in the $e \tau$ plane, there
remains a nonzero transition moment between the
heavy and light mass eigenstates
\begin{equation}
\mu = \mu_{e \mu} \cos \theta + \mu_{ \mu \tau} \sin \theta
\end{equation}
In terms of the neutrino mass it can be given by
\begin{equation}
\mu \simeq 2 e \frac{m_{DM}}{M_1 } \sin\theta F(m_i, M_1, M_2)
\end{equation}
where $F$ is a smooth function of the internal charged
lepton masses $m_i$ and Higgs masses $M_1$ and $M_2$.
varying from 1 to 10 for reasonable values of the parameters.
Within the experimental limits on neutrino mixing and
Higgs boson masses that follow from negative \neu
oscillation \cite{PDG92} and Higgs boson searches
\cite{Aleph92} one can reach values up to $3\times 10^{-14} \mu_B$
for $m_{DM} = 30$ eV.

In order to account for the photon flux ionizing interstellar hydrogen
one should have $M \sim 30$ eV and $\mu \sim 10^{-14} \mu_B$.
It is easy to verify that both requirements can easily be met in our model.
However, a 30 eV hot dark matter neutrino
made out of two active components, \nm and \nt,
may cause problems with several observations.
For example, the ratio of the neutrino density
to the critical density is determined from
\begin{equation}
\Omega_\nu = \frac{\sum m_\nu }{h^2 91 \unit{eV} },
\end{equation}
where $h$ is the (dimensionless) Hubble constant.
If we require that the density of the universe be
the critical density, as suggested by inflation,
the age of the universe is given by
\begin{equation}
t= \frac{2 }{3 H}.
\end{equation}
Two 30 eV neutrinos would then imply that the age
of the universe to be 8 Ga, which contradicts the
observational lower limit 10 Ga. In principle this
age problem could be avoided by relaxing the requirement
that $\Omega = 1$. One could have acceptable ages of more
than ten billion years with $\Omega \gsim 3$.

Models with only hot dark matter have also difficulties to
explain the structures observed in the universe. These could
be corrected by imposing strong bias on these schemes, as could
arise from topological defects in the very early universe,
e.g. cosmic strings.
These could produce substantial seeds
for galaxy formation while the galactic halos can be explained
by baryonic matter. In fact there seem to be some evidence
that HDM plus cosmic strings has roughly the properties that
seem to be required for galaxy formation and is in agreement
with COBE results \cite{AlbrechtStebbins92}.

\section{Rare Decays}

Our choice of the symmetry group allows tau leptons to decay to final states
involving electron and something else. With the couplings defined in the
lagrangian, the fastest new decay mode would be the decay to two electrons
and a positron. It occurs through the tree-level graph of Figure 2,
and its branching ratio is given by
\beq
BR (\tau \ra 3 e ) \simeq
                \frac{g_{e\tau}^2 g_{ee}^2}{g_2^4}
                \frac{M_W^4}{M_1^4}.
\eeq
The experimental constraints of our model allow this to be
as large as the present experimental limit ($ 4 \times 10^{-5}$) \cite{PDG92},
which makes it reasonable to search for this process in
future tau factories. The most obvious choice of the parameters providing the
solutions to dark matter, solar neutrino and atmospheric neutrino problems
yields branching ratios inaccessible to any feasible experiment, but there is
a certain completely natural region of parameters giving observable rates
while at the same time reconciling the above problems.
For example, our previous sample choice
gives $BR(\tau \to 3e) \sim 3 \times 10^{-6}$.

At one-loop level one can have radiative tau decays. For a certain range of
parameters this can also be observable, but it is likely to be smaller than
the decay to electrons. The branching ratio is given by
\beq
BR (\tau \ra e + \gamma) \simeq 4 \times 10^5
                {g_{e\tau}^2 g_{\tau\tau}^2}\left(\frac{\rm
GeV}{M}\right)^4.
\eeq

If we choose to break the symmetry spontaneously, the tau can
decay into an electron and a majoron, with a branching ratio
\beq
BR (\tau \ra e + J) \simeq 5 \times 10^6
                g_{e\tau}^2 g_{\tau\tau}^2
                \frac{M_\kappa^4 \VEV{\sigma}^2 }{M^8}
\eeq
where $M_\kappa$ is the cubic $\kappa \Xi \sigma$
coupling constant and all masses are expressed in GeV.
For the above parameter choice and similarly for any
other natural choice that reconciles all the three
problems this is expected to be below experimental
detectability. However, relaxing the atmospheric neutrino condition
the majoronic decay could be as large as the current search limit.

All other rare decays, especially those involving muons, are more
suppressed, as they require the violation of our global symmetry.

\section{The $e-\tau+\mu$ Model}

An entirely similar, physically inequivalent model
can be obtained by replacing the underlying global
\sym $e-\mu+\tau$ by the $e-\tau+\mu$ symmetry.
For this case we can apply the previous formulae,
with $\mu$ and $\tau$ interchanged. A very important
difference is that now the most relevant mixing angle
is that between electrons and muons, which is given by
\begin{equation}
 \tan \theta \simeq \frac{m}{M} \simeq
\frac{f_{\tau e}}{f_{\mu \tau}}
\end{equation}
which controls the rates for \nm to \ne oscillations
at accelerators. Existing laboratory experiments
restrict this mixing angle to $\sin^2 2\theta < 0.004$ \cite{PDG92},
which gives for the coupling constants a bound
$f_{e\tau} < 0.03 f_{\mu \tau}$.

One can easily generate a mass matrix that satisfies our previous dark
matter, solar and atmospheric neutrino conditions, for example with the choice
$f_{e\tau}= 10^{-5} $, $f_{\mu\tau}= 4 \times 10^{-4} $ ,
$g_{ee}= 6 \times 10^{-3}$,
$g_{\mu e} = 3 \times 10^{-4} $, $g_{\mu\mu}=0.2$,
$\xi_e=10^{-5} $, $\xi_\mu= 2 \times 10^{-5}$, $M_{s\phi\tilde{\phi}}=5 $
GeV, $ M_{\kappa\Xi}=100$ GeV,
$M_{s\eta\kappa}= M_{\eta\eta\Xi}=170$ GeV,
$M_0= 160 $ GeV, $v = 80$ GeV, $\tilde{v}=240$ GeV, $\sin \beta = 0.2$.
On the other hand, it is more difficult to generate sizeable magnetic
transition moments, because of the more stringent bound for the
mixing of electron and muon neutrinos, and less favourable integral factors.

The estimated branching ratios of lepton flavour violating
muon decays and their present experimental limits \cite{PDG92}
are given by
\begin{eqnarray}
BR (\mu \ra 3 e ) &\simeq&
                \frac{g_{e\mu}^2 g_{ee}^2}{g_2^4}
\frac{M_W^4}{M_1^4} \leq 10^{-12}\\
BR (\mu \ra e + \gamma) &\simeq& 4 \times 10^5
                {g_{e\mu}^2 g_{\mu\mu}^2}
\left(\frac{\rm GeV}{M}\right)^4 \leq 5 \times 10^{-11}.  \\
BR (\mu \ra e + J) &\simeq& 10^9
                g_{e\mu}^2 g_{\mu\mu}^2
                \frac{M_\kappa^4 \VEV{\sigma}^2 \rm{GeV}^2 }{M^8}
\leq 2.6 \times 10^{-6},
\end{eqnarray}
where the last process exists only if we have the spontaneous
violation of the underlying global symmetry.
One can easily verify that all these rates can reach
values within the sensitivities of present and future
experiments. Our  sample set of values gives $\sim 4 \times 10^{-13}$
for the case of $\mu \ra 3e$, $\sim 2 \times 10^{-12}$ for
$ \mu \ra e\gamma$ decays and $\sim 2 \times 10^{-6}$ for
the majoron emitting decay mode (choosing $\VEV{\sigma}=170$ GeV).
It is quite possible to obtain values of these branching ratios
in excess of present limits \cite{PDG92}!

\section{Discussion}

We have analysed interesting symmetries
of the leptonic electroweak interaction which are
suggested in order to reconcile the solar
and atmospheric neutrino data with the
hot dark matter component in the universe.
This requires the presence of a fourth light
\neu species \ns, singlet under the \21 symmetry.
We have also provided concrete models that embody
this symmetry and in which the HDM is a quasi-Dirac
neutrino formed by \nm and \nt whose mass $m_{DM}$ is
induced at the one-loop level. Unlike previous
suggestion, in the present model the sterile
\neu is ultralight, and the solar \neu deficit
is explained via nonadiabatic conversions
of \ne to \ns, while the atmospheric \neus are
explained via maximal \nm to \nt oscillations.
Both oscillations are generated as breaking
effects and arise as higher order loop
diagrams which can naturally provide the required
small mass splittings.

For $m_{DM} \sim 30$ eV the radiative
HDM decay leads to photons that can
ionize interstellar hydrogen. The
 price is that, for $\Omega=1$,
we have a relatively young universe
with 8 Ga. The complete inner
consistency of this possibility
from the particle physics point
of view and its ability to explain
other astrophysical and cosmological
puzzles in an attractive way highlight,
in our opinion, the interest in further
pushing on the determination of the age
parameter.

The $\nu_e$ to $\nu_\tau$ oscillations
can have experimentally observable rates,
with no appreciable \nm oscillations expected
at accelerator experiments.
In addition, this model can lead to tau number violating processes
\sa $\tau \ra e + \gamma$, $\tau \ra e + J$ and $\tau \ra 3e$
with rates that can lie within the sensitivities of future
tau factories.

In the variant model described in section 6
the oscillations that are phenomenologically
relevant for accelerator experiments are
$\nu_e$ to $\nu_\mu$, and they can have
experimentally observable rates. In this case
the lepton flavour violating processes
involve muons, e.g.
$\mu \ra 3e$,
$\mu \ra e + \gamma$ and
$\mu \ra e + J$. Their rates can lie within the
sensitivities of present experiments.

We stress that these models are compatible with laboratory,
astrophysical and cosmological observations,
including primordial nucleosynthesis limits.
The new Higgs bosons present in these models also
modify the baryogenesis conditions \cite{BGLAST}.
Conventional models for the generation of the baryon asymmetry
within the standard model are likely to contradict \cite{Shap} the laboratory
limits for the Higgs boson masses \cite{Aleph92}. These constraints may be
avoided in multi-Higgs models, like ours. We can also have large
CP-violation either on the Yukawa couplings, or in the Higgs sector.
These effects would not be restricted by the physics of the quark
sector, as our new Higgses do not couple to quarks.

{\large \bf Acknowledgement}

This work was supported by the
Spanish Ministry for Education and Science.
We thank David Caldwell, Vadim Kuzmin, Anjan
Joshipura and Daniele Tommasini for
discussions; Adrian Mellott and Dennis
Sciama for bringing to our attention the
literature on unstable dark matter and
the ionization of interstellar hydrogen.

\clearpage

\newpage
\section*{Figure Captions}
\noindent
{Fig. 1}\\
Diagrams generating neutrino masses from
radiative corrections (here $\ell ,\ell^\prime = e, \tau$).
The blobs denote explicit $e-\mu+\tau$ soft brweaking terms.
\\
\noi
{Fig. 2}\\
Diagrams responsible for the decays
$\tau \ra e + J$ and $\tau \ra e + \gamma$.
\newpage

\begin{table}
\begin{center}
\begin{math}
\begin{array}{|c|crrr|} \hline
 &  T_3 & Y & G & \\
\hline
\Psi_{L\ell}  &  \frac{1}{2}          & -1 & 1 & \\
\Psi_{L\mu}  &  \frac{1}{2}          & -1 & -1 & \\
\ell_{R}   &  0 & -2 & 1 & \\
\mu_{R}   &  0 & -2 & -1 & \\
\nu_s &  0 & 0 & -\frac{1}{2} & \\[0.2cm]
\hline
\phi &  \frac{1}{2}          & 1 & 0 & \\
\tilde{\phi} &  \frac{1}{2}  & 1 & 0 & \\
\hline
s^- & 0 & -2 & 0 & \\
\eta^- & 0 & -2 & \frac{3}{2} & \\
\hline
\kappa^{--} & 0 & -4 & 2 & \\
\Xi^{--} & 0 & -4 & \frac{5}{2} & \\
\hline
\sigma & 0 & 0 & \frac{1}{2} & \\
\hline
\end{array}
\end{math}
\end{center}
\caption{$SU(2) \ot U(1)_Y \ot U(1)_G$ assignments of the
leptons and Higgs scalars in the model of section 3.
Another model was discussed in section 6. In addition,
for each case one can obtain variant models by exchanging
the signs of the $U(1)_G$ charges of
$\nu_s$, $\eta$, $\kappa$, $\Xi$ and $\sigma$. Both models
can be formulated with explicit or spontaneous breaking of
the $U(1)_G$ symmetry. Only in the latter case the scalar
$\sigma$ is present.}
\end{table}
\end{document}